\documentclass[reprint,prd,twocolumn,superscriptaddress,showpacs,nofootinbib,floatfix,preprintnumbers]{revtex4-1}
\usepackage{graphicx, bm}

\def\mET{E_T \hspace{-1.2em}/~~}

\begin{document}
\preprint{MAN/HEP/2013/09}
\title{Probing Heavy-Light Neutrino Mixing  in Left-Right Seesaw Models at the LHC}

\author{Chien-Yi Chen}
\affiliation{Department of Physics, Brookhaven National Laboratory, Upton, New York 11973, USA}

\author{P. S. Bhupal Dev}
\affiliation{Consortium for Fundamental Physics, School of Physics and Astronomy, University of Manchester, Manchester M13 9PL, United Kingdom}

\author{R. N. Mohapatra}
\affiliation{Maryland Center for Fundamental Physics and Department of Physics, University of Maryland, College Park, Maryland 20742, USA}

\begin{abstract}
We show that in TeV-scale left-right (L-R) symmetric seesaw models, there are new dominant 
contributions to the collider signals of heavy Majorana neutrinos arising from the heavy-light neutrino mixing, which directly probe the seesaw matrix in a certain class of models. 
We propose a way to distinguish this contribution from the widely discussed one that only probes the Majorana nature of the heavy right-handed neutrinos,
 by analyzing some simple kinematical variables. We find that in this class of L-R seesaw models the existing LHC data already yield slightly stronger constraints on the heavy-light neutrino mixing than those derived 
for standard seesaw models, and the improvement will be significant as more data are collected.
\end{abstract}

\maketitle

\section{Introduction}
The neutrino oscillation data unambiguously establish that neutrinos have tiny but non-zero masses, the explanation of which calls for physics beyond the Standard Model (SM). A simple paradigm for understanding the smallness of left-handed (LH) neutrino masses is the (type-I) seesaw mechanism~\cite{type1} where one introduces a set of heavy SM singlet Majorana fermions $N$ breaking the $(B-L)$-symmetry. The seesaw matrix has the generic form in the $ (\nu_L, N)$ space:
\begin{eqnarray}
\left(\begin{array}{cc}0 & m_D\\ m^{\sf T}_D & M_N\end{array}\right) 
\end{eqnarray}
where $m_D$ is the Dirac mass term which mixes the $\nu$ and $N$ states, and $M_N$ is the Majorana mass term for $N$. This leads to the seesaw formula for light neutrinos of the form~\cite{type1}
\begin{eqnarray}
M_\nu \simeq -m_DM_N^{-1}m_D^{\sf T}, 
\label{eq:1}
\end{eqnarray}
and a heavy-light neutrino mixing of order $m_DM_N^{-1}$~\cite{valle}. Thus there are two key aspects to the seesaw mechanism: the Majorana mass of the heavy neutrino, and the mixing between the heavy and light neutrinos. To probe the seesaw paradigm experimentally, one must therefore test both the Majorana nature of $N$ and the heavy-light neutrino mixing effects. There are two possible ways to do this. The first well known way is to test for the Majorana nature of both the heavy and light neutrino masses via searches for the neutrinoless double beta decay ($0\nu\beta\beta$) and disentangle the heavy neutrino effect~\cite{0v2breview} which however does not 
necessarily probe the heavy-light neutrino mixing. The second way is to directly look for the presence of 
heavy-light mixing, which can manifest in several ways, e.g., (i) via departures from unitarity of the PMNS 
neutrino mixing matrix~\cite{unitarity}, which can be probed in neutrino oscillation experiments as well as lepton flavor violation (LFV) searches, and (ii) via their signatures in colliders~\cite{Drewes:2013gca}. Clearly for these latter tests of seesaw to be effective, the mixing parameter $m_DM_N^{-1}\equiv V_{\ell N}$ must be significant and this requires that $M_N$ must be small (in the TeV range) and $m_D$ large (in the few GeV range) simultaneously. It is the second aspect of testing seesaw at colliders that we focus on in this 
paper.

To proceed with details, we remind the reader that the simplest implementation of the seesaw paradigm is to add the gauge-singlet neutrino field $N$ with a Majorana mass $M_N$ to the SM. The seesaw scale (synonymous with $M_N$) then remains an adhoc parameter unconnected to any new physics or symmetry. We will call this scenario the SM-seesaw in what follows. 
On the other hand,
these heavy neutrinos $N$ naturally arise as the right-handed (RH) partners of the LH neutrinos in the Left-Right (L-R) symmetric extension of the SM which was originally introduced~\cite{LR} in order to understand the origin of parity violation in weak interactions at low energies. The minimal L-R symmetric theory, based on the $SU(3)_c\times SU(2)_L\times SU(2)_R\times U(1)_{B-L}$ gauge group, provides a natural explanation of the seesaw scale as connected to the $SU(2)_R\times U(1)_{B-L}$ breaking scale. The smallness of the LH neutrino mass in these theories is connected to the extent to which the RH-current effects in weak interactions are suppressed at low energy.  We will call this scenario the L-R seesaw. Thus, a TeV-scale L-R symmetric theory provides an attractive class of seesaw models that can be probed at the LHC~\cite{goran}. 

\begin{figure*}[htb]
\centering
\begin{tabular}{cccc}
\includegraphics[width=4cm]{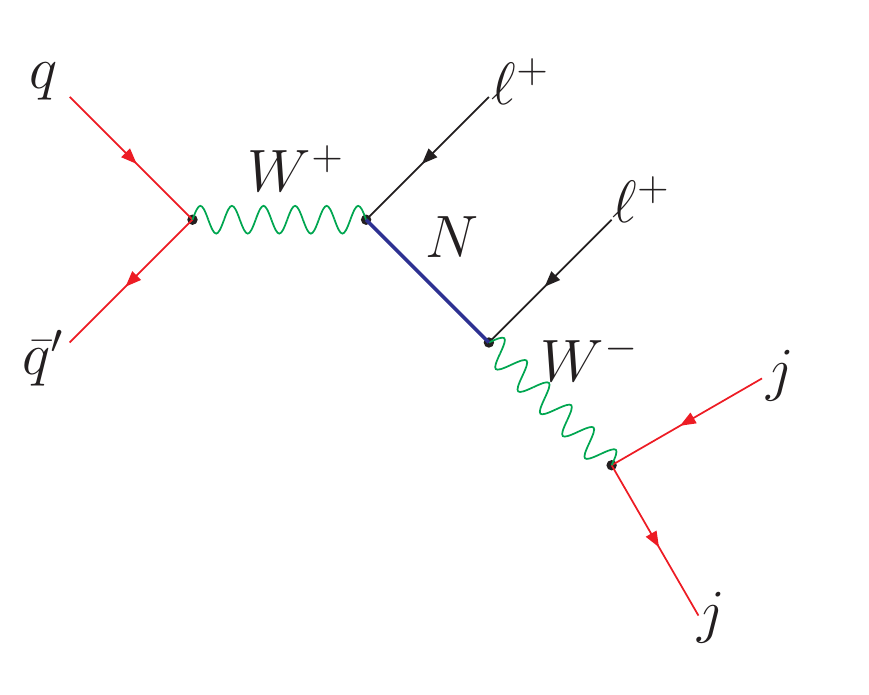} & 
\includegraphics[width=4cm]{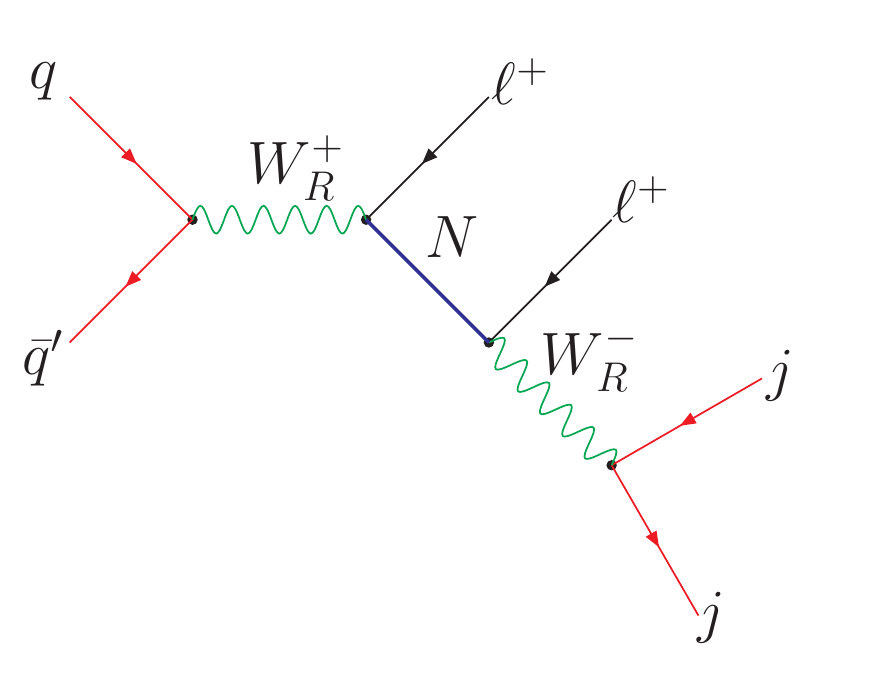} &
\includegraphics[width=4cm]{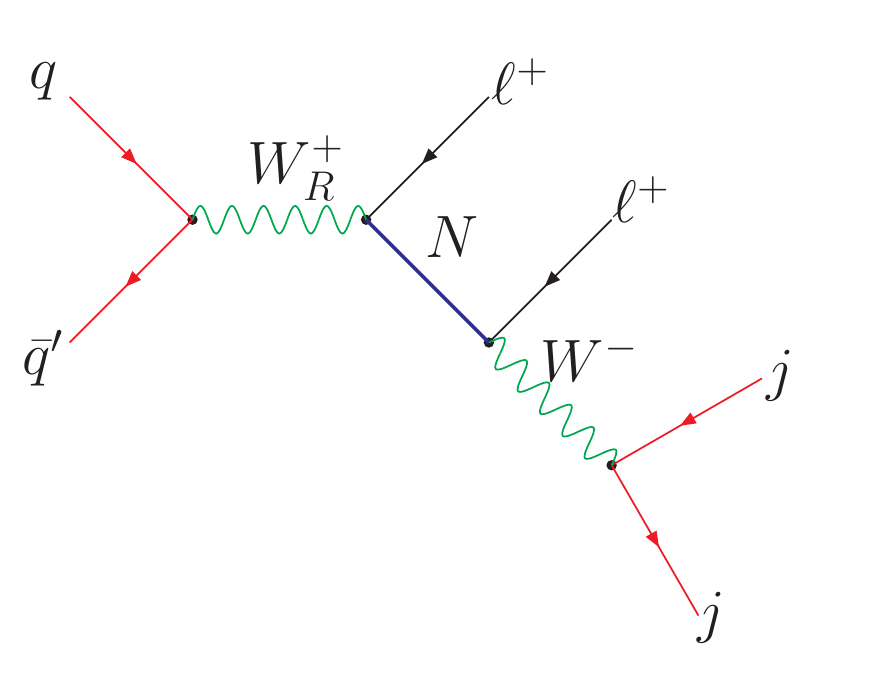} &
\includegraphics[width=4cm]{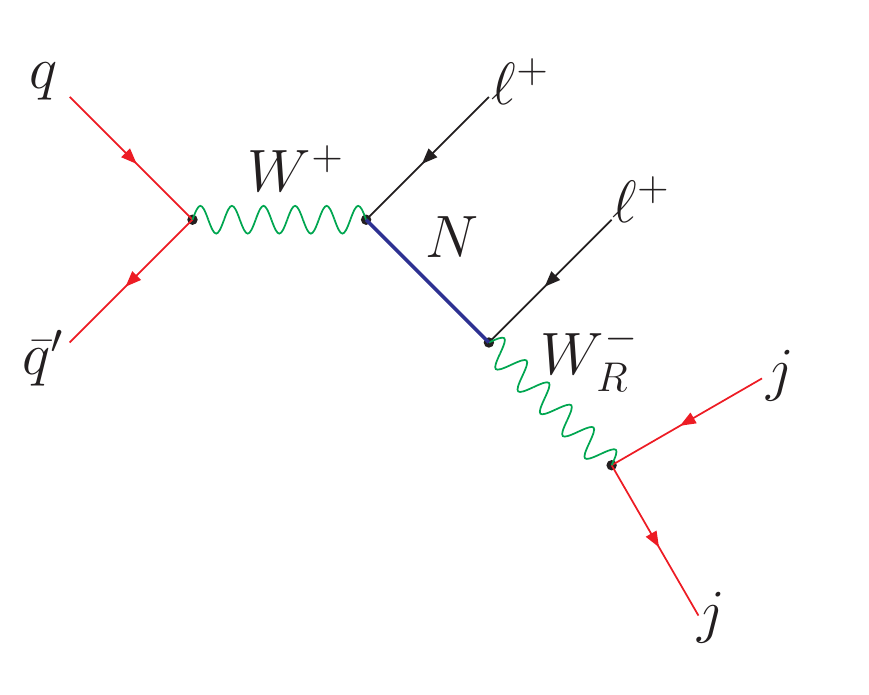} \\
(a) $LL$ & (b) $RR$ & (c) $RL$ & (d) $LR$ 
\end{tabular}
\caption{The Feynman diagrams contributing to the `smoking gun' collider 
signal of seesaw %$\ell^\pm\ell^\pm jj$ of a heavy Majorana neutrino 
in the minimal L-R model. }
\label{fig1}
\end{figure*}

As noted, for the case of SM-seesaw,  the Majorana mass $M_N$  is hard to test in colliders without the help of the heavy-light neutrino mixing $V_{\ell N}$. The full 
seesaw mechanism can then manifest itself as final states with same-sign dileptons 
plus two jets without missing energy ($\ell^\pm\ell^\pm jj$), arising from the Feynman diagram shown in 
Fig.~\ref{fig1}a. 
This signal depends crucially on the heavy-light mixing and can effectively probe the heavy neutrino masses $M_N$ only up to a few hundred GeV as has been extensively discussed in the literature~\cite{theory-LL}. It must be stressed that any positive signal would not only signify the Majorana character of the heavy sub-TeV neutrino  $N$ but also a specific  non-generic structure of $m_D$. The reason is that in generic (``vanilla'') seesaw case, we expect the heavy-light mixing $V_{\ell N}\sim \sqrt{m_\nu/M_N}$ which is very tiny for TeV-scale $M_N$ due to the smallness of the light neutrino masses (the current upper limit on $m_\nu\leq 0.1$ eV~\cite{Planck}), making the collider signal unobservable.  Only if the Dirac matrix $m_D$ in Eq.~(\ref{eq:1}) has specific forms (see e.g.,~\cite{cancel, Mitra:2011qr}) can $V_{\ell N}$ be significant enough to have observable  lepton number violation (LNV) at the LHC~\cite{theory-LL}. The latter can reveal underlying symmetries of the lepton sector, which will be an important step towards a full understanding of the neutrino mass physics. 
We note parenthetically that the other manifestation of LNV, namely $0\nu\beta\beta$, 
receives dominant 
contribution only from the light neutrino mass in this case~\cite{Blennow:2010th} (except when the light neutrino contribution vanishes due to cancellation~\cite{Mitra:2011qr, Ibarra:2011xn}). 

However, in the L-R symmetric embedding of TeV-scale seesaw, 
%the situation changes completely. as has been already recognized~\cite{KS}. 
the presence of RH gauge interactions lend considerable richness to the manifestations of seesaw in experiments~\cite{Tello:2010am}. Not only are there new contributions to $0\nu\beta\beta$ from RH gauge bosons ($W_R$)~\cite{MS}, but the 
profile of seesaw manifestation at colliders changes dramatically~\cite{KS}. In fact, the $\ell^\pm\ell^\pm jj$ signal now receives three new contributions from different combinations of $W_R$ exchange and heavy-light neutrino mixing (Figs.~\ref{fig1}b-\ref{fig1}d). The contribution which arises from the exchange of two $W_R$ bosons is the one that has been widely discussed for the L-R seesaw case~\cite{theory-RR}. However for certain specific textures of Dirac mass matrix, which lead to an enhanced heavy-light neutrino mixing, the profile of the ``smoking gun''  $\ell^\pm\ell^\pm jj$ signal changes drastically. The goal of this paper is to explore the relative magnitude of the heavy-light mixing contribution compared to the $W_RW_R$ contribution at the LHC and assess their impact on our understanding of the seesaw paradigm.
 
Important for the collider discussion are the relative values of the $W_R$ and $N$ masses. There are theoretical arguments based on vacuum stability~\cite{rnm} which suggest that the heavy neutrinos in the minimal L-R seesaw model are 
lighter than the RH gauge bosons for a large range of parameters. 
We will therefore consider this mass ordering $M_N<M_{W_R}$ in this paper 
(although going beyond the minimal version, one could avoid this restriction). 
A major implication of this, as shown in this paper, is that for RH gauge boson masses below 4-5 TeV, 
when it can be produced at the $\sqrt s$=14 TeV LHC with a decent cross 
section, its decay to the on-shell heavy RH neutrinos will allow a new probe of its mixing with the light neutrinos for a wider heavy neutrino 
mass range of up to a few TeVs from a study of $\ell^\pm\ell^\pm jj$ final 
states. This information, together with the light neutrino mixing parameters extracted from neutrino 
oscillation data, should suffice to fully 
determine the Dirac mass matrix in the minimal L-R model, and hence, facilitate its testability in other low energy experiments.
%%%%%%%%%%%%%%

%This paper is organized as follows: in sec. 2, we present some examples of Dirac textures that lead to enhanced heavy-light mixing with TeV scale seesaw; in sec. 3, we present the phase diagram for regions of heavy light mixing and $W_R$ mass where the former makes dominant contribution to the $\ell^\pm\ell^\pm jj$ signal; in sec. 4, we apply this discussion to obtain limits on the heavy light mixing for this case and point out future prospects for improvement; in sec.5, we point out ways to distinguish between the heavy-light mixing contribution from $W_RW_R$ contribution in colliders and in sec. 6, we present our concluding remarks.
%%%%%%%%%%%%%%
\section{Textures with 
enhanced $V_{\ell N}$ in TeV-seesaw}
%%%%%%%%%%%%%%%
As is well known and also as emphasized in the introduction, for generic forms of both the Dirac mass matrix $m_D$ and the RH neutrino mass matrix $M_N$, the seesaw formula in Eq.~(\ref{eq:1}) implies that the heavy-light mixing parameter $V_{\ell N} \simeq \sqrt{m_\nu/M_N}$ which is a tiny number regardless of whether the seesaw scale is in the TeV range or higher. This keeps its effect shielded from being probed by either collider or low energy experiments. However, there are some special textures for $m_D$ for which even with TeV-scale seesaw, the mixing parameter $V_{\ell N}$ can be significantly enhanced whereas the neutrino masses remain naturally small. We present only one example here to illustrate our case, although several others have been discussed in the literature~\cite{cancel, Mitra:2011qr}. Consider the matrices $m_D$ and $M_N$ of the following form:
\begin{eqnarray}
m_D=\left(\begin{array}{ccc} a & \delta_1 & \epsilon_1\\ b & \delta_2 & \epsilon_2\\ c &\delta_3 & \epsilon_3\end{array}\right)~{\rm and}~% \\ \nonumber
M_N=\left(\begin{array}{ccc} 0 & M_1 & 0\\M_1 & \delta M & 0 \\ 0&0&M_2\end{array}\right)
\label{eq:texture}
\end{eqnarray}
with $\epsilon_i, \delta_i \ll a,b,c $ and $\delta M \ll M_i$. In the limit of $\epsilon_i, \delta_i, \delta M\to 0$, the neutrino masses vanish, although the heavy-light mixing given by $V_{\ell N_i}=m/M_i$ (with $m=a,b,c$) can be quite large. The neutrino masses given by the seesaw formula become proportional to products of  $\epsilon_i$ and $\delta_i$. If by some symmetry one can guarantee the smallness of $\delta_i$ and $\epsilon_i$, then  we have a TeV scale seesaw model with enhanced $V_{\ell N}$. These mass textures can be embedded into L-R models~\cite{dlm} and will have other phenomenological implications, e.g. ``large'' LFV, violation of unitarity of the PMNS mixing matrix, etc. It is the impact of these scenarios in colliders which is the main focus of the rest of this paper. Note that while we have presented only one example of such non-generic Dirac mass matrix in Eq.~(\ref{eq:texture}), our following results are also applicable to other Dirac textures discussed in the literature.
\section{The Left-Right Phase Diagram}
%%%%%%%%%%%%%
In this section, we present the regions of heavy-light mixing parameter and RH gauge boson masses where the 
mixing effects will provide the dominant contribution to the $\ell^\pm\ell^\pm jj$ signal. Clearly there will be flavor dependence in this signal, depending on the underlying Dirac mass texture; we do not discuss 
those details here and show our results for a generic case.
There are four classes of Feynman diagrams in the minimal L-R model which can 
lead to the $\ell^\pm\ell^\pm jj$ final states (Fig. 1). We denote these 
diagrams as (a) $LL$, (b) $RR$, (c) $RL$, and (d) $LR$, according to the chirality of 
the final state lepton-pair. The most widely studied of these are the $LL$ and 
$RR$ diagrams -- the first one in the context of SM-seesaw~\cite{theory-LL} and the second one in LR models~\cite{KS,theory-RR}.
The channel in Fig 1a is a clear probe of the seesaw matrix in both SM-seesaw 
and L-R seesaw models, but its effectiveness solely relies on the heavy-light mixing $|V_{\ell N}|^2$, and is limited to $M_N$ only up to a 
few hundred GeV. Experimentally, the mass range $M_N=100$ - 300 GeV has been 
explored at the LHC for $\ell=e,\mu$~\cite{CMS-LL, ATLAS-LL}, 
and direct upper limits on 
$|V_{\ell N}|^2$ of the order of $10^{-2}$ - $10^{-1}$ have been set. 
We note here that the complementary limits from electroweak 
precision tests and lepton flavor violating processes are roughly one to two 
orders of magnitude stronger (for a review, see Ref.~\cite{Atre:2009rg}).   

In case of  Fig 1b, the heavy neutrinos are produced on-shell via the decay of 
an RH gauge boson and they then subsequently decay into a three-body final state via an off-shell $W_R$. This diagram gives the 
dominant contribution if the 
heavy-light mixing is assumed to be very small which is of course the naive 
expectation in the ``vanilla'' type I seesaw case as noted above. Using this 
channel, LHC exclusion limits are derived in the ($M_N,M_{W_R}$) 
plane~\cite{ATLAS-RR, CMS-RR}, and currently exclude $M_{W_R}$ up to 2.5 TeV 
for a TeV-scale $M_N$. Note that these limits are independent of the Dirac 
neutrino Yukawa coupling characterizing the heavy-light mixing, and therefore, 
do not probe the seesaw matrix. 

The contributions shown in Figs.~\ref{fig1}c and \ref{fig1}d, on the other 
hand,  
necessarily involve the heavy-light neutrino mixing.\footnote{The heavy-light mixing also contributes to 
$0\nu\beta\beta$ in L-R models~\cite{rnm,Hirsch:1996qw}. 
Again these effects are small for generic seesaw matrix, but could be important for large mixing~\cite{Nemevsek:2012iq}.}  In fact, the $RL$ diagram could give the dominant 
contribution to the $\ell^\pm\ell^\pm jj$ signal if the mixing $|V_{\ell N}|$ is non-negligible 
and/or the $W_R$ gauge boson is not too heavy. There are two reasons for this 
dominance: (i) this contribution leads to a production rate $\sigma(pp\to W_R\to N\ell^\pm)$ which is independent of mixing  and only suppressed by 
$(M_{W}/M_{W_R})^4$ (as in the $RR$ case), and can therefore dominate over the 
$LL$ contribution which depends on $|V_{\ell N}|^2$; (ii) the decay of the heavy 
neutrino in this case is no longer suppressed by the phase space, 
since it can have a 
two-body decay via on-shell $W$: $N\to \ell^\pm W\to \ell^\pm jj$ (as in the $LL$ case). Hence, for a sizable range of the mixing and RH gauge boson mass, 
the RL mode is expected to be dominant for the heavy neutrino signal $\ell^\pm\ell^\pm jj$ at the LHC and could constitute a clear probe of the 
seesaw matrix. It is surprising that this contribution has not been taken into account in the collider analyses so far, although the importance of this 
contribution has been discussed sporadically, e.g., in the context of a 
comparative study between heavy Majorana and Dirac 
neutrinos~\cite{Chen:2011hc}, and in determining the chirality of the heavy gauge boson~\cite{Han:2012vk}. 

The remaining possibility, namely, the $LR$ contribution (Fig.~\ref{fig1}d) is 
doubly suppressed by the mixing as well as phase space, and hence, always 
smaller than at least two of the other three contributions discussed above. Hence, we will not analyze this diagram in details in what follows.

\begin{figure}[tb]
%\vspace{0.2cm}
\centering
\includegraphics[width=4.00cm]{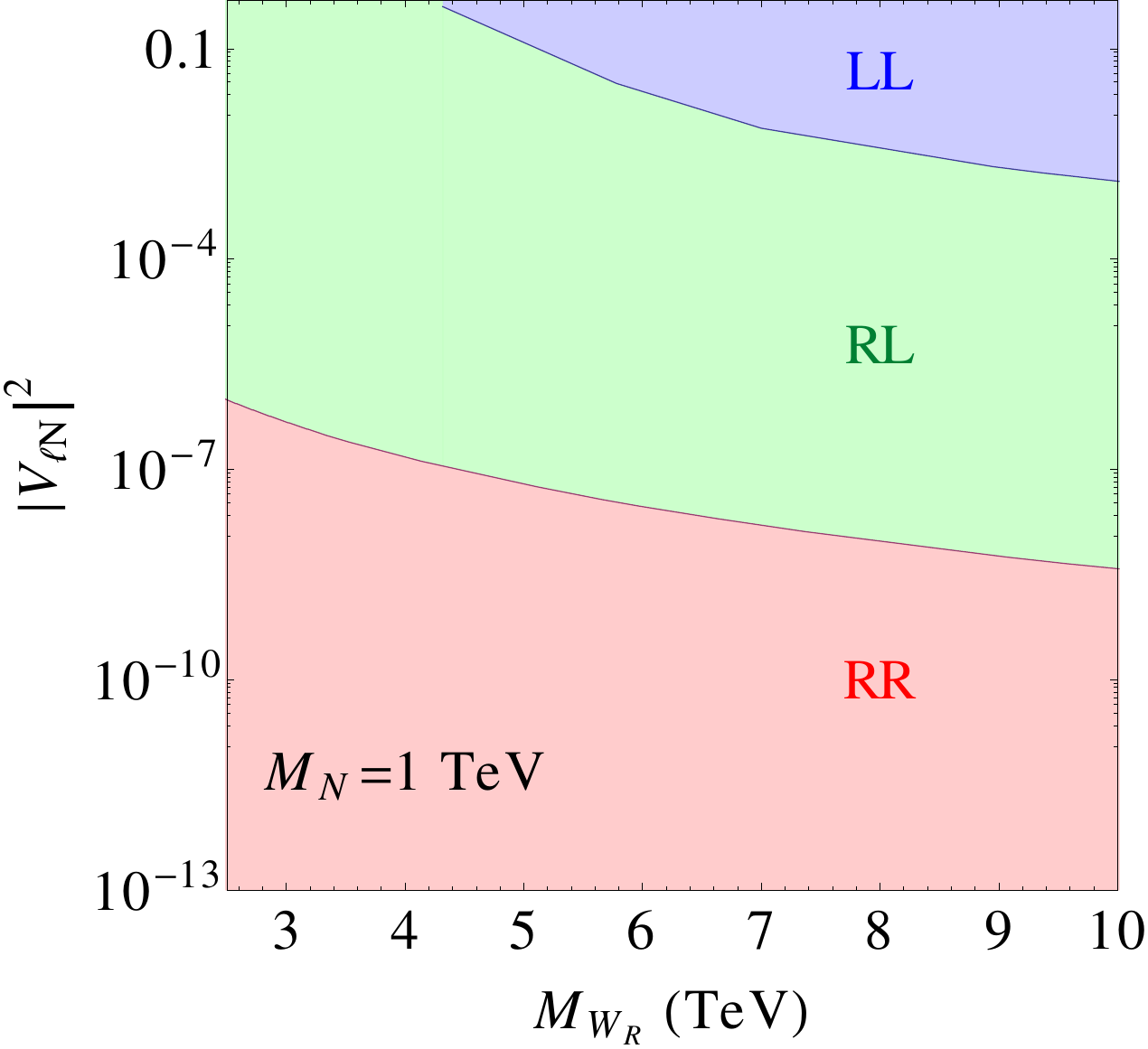}
%\hspace{0.0cm}
\includegraphics[width=4.00cm]{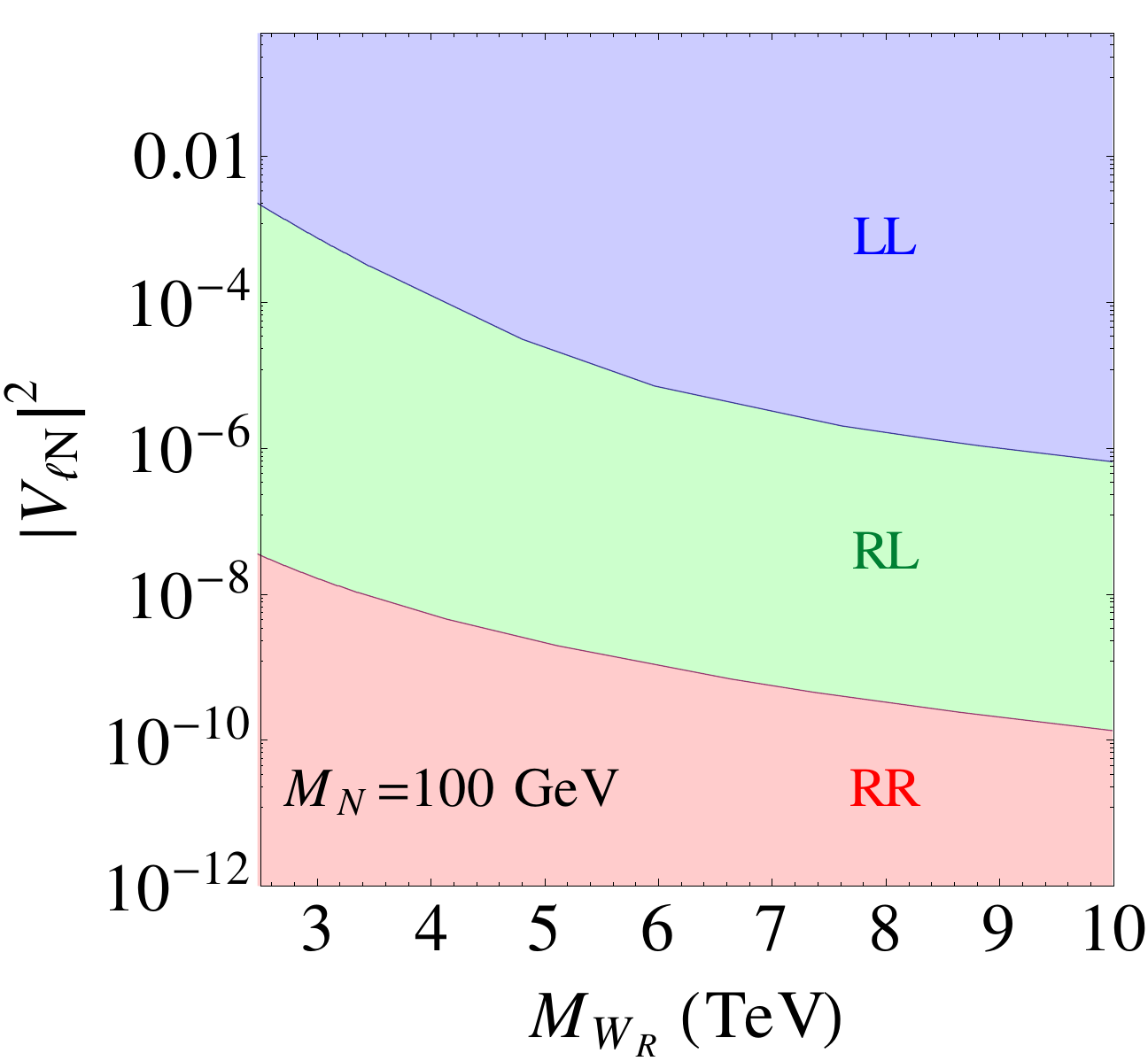}
\caption{Phase diagram for the minimal L-R seesaw model.}%a heavy Majorana neutrino in the minimal LR model.}
\label{fig2}
\end{figure}

The regions of dominance for various contributions discussed above are 
shown in Fig.~\ref{fig2} (we call this the ``L-R phase diagram") for two 
typical choices of the  heavy neutrino mass $M_N=100$ GeV and 1 TeV. The upper (blue) shaded region with large mixing is where the LL contribution to the 
$\ell^\pm\ell^\pm jj$ signal is dominant, whereas the lower (red) shaded 
region with small mixing is dominated by the RR contribution. The middle 
(green) region is where the RL contribution is dominant and it clearly spans a 
wide parameter space of the model. In particular, it can probe the seesaw mixing all the way down to $|V_{\ell N}|^2\geq 10^{-8}$, close to the 
``vanilla" seesaw expectation of $m_\nu/M_N$. 
%Thus combining collider limits 
%with fitting the neutrino oscillation data using Eq.~(\ref{eq:1}) which also depends on heavy-light mixing and heavy neutrino mass scale can play a decisive role in testing TeV-scale LR seesaw models.

\begin{figure}[htb]
\centering
\includegraphics[width=4cm]{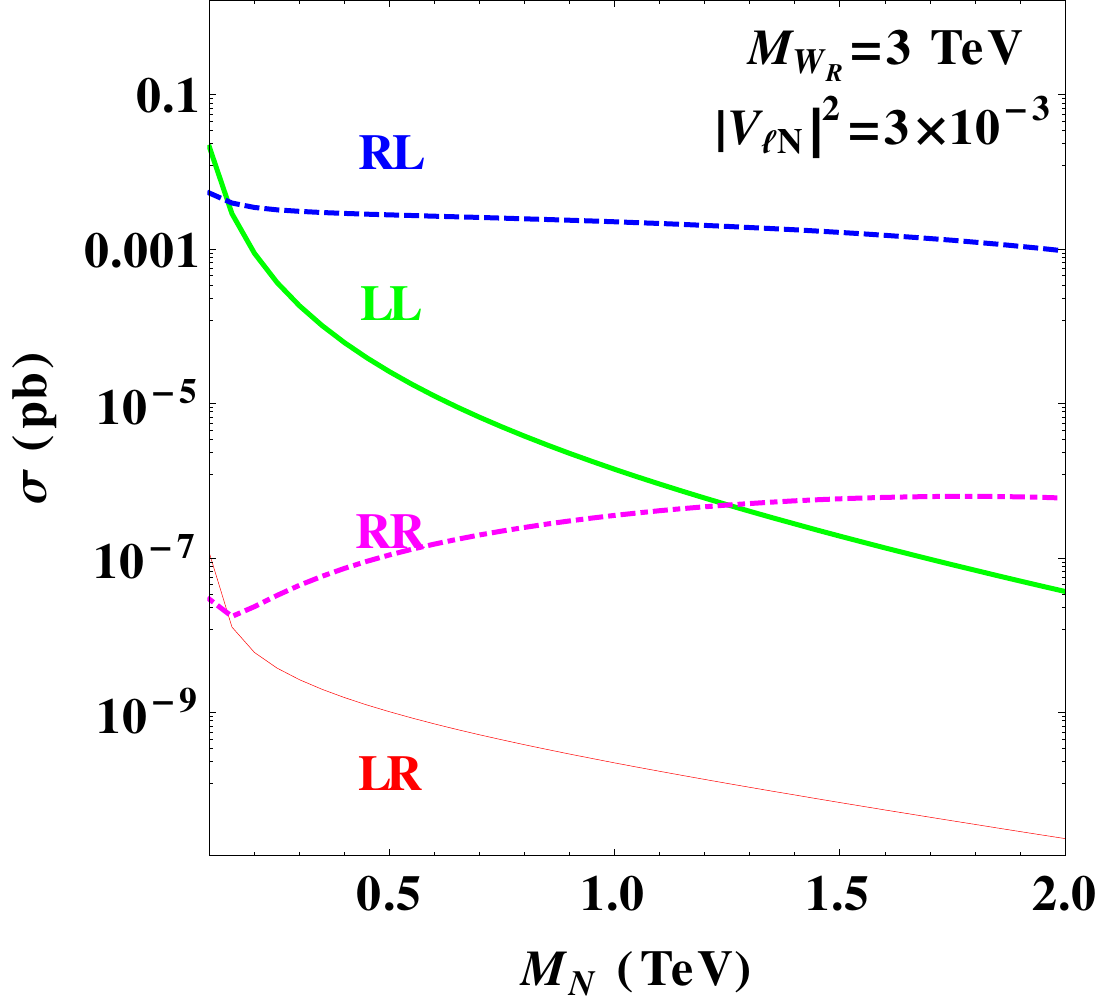}
%\hspace{0.0cm}
\includegraphics[width=4cm]{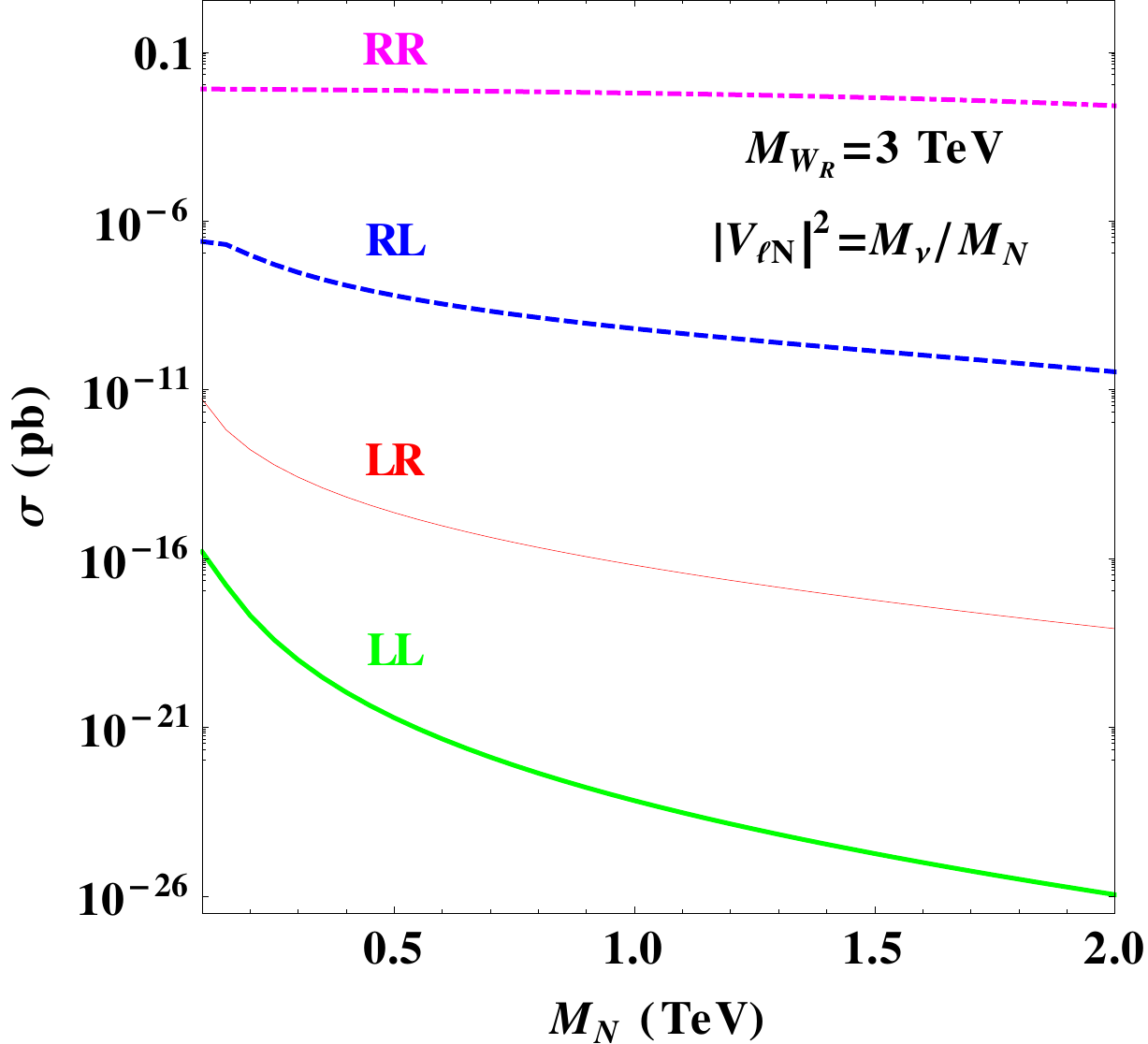}
\caption{Comparison of the signal cross sections for various modes shown in Fig.~\ref{fig1} for two benchmark scenarios.}
\label{fig3}
\end{figure}

To further illustrate our point, we compare the magnitudes of signal 
cross section for the processes shown in Fig.~\ref{fig1} as a function of the RH neutrino mass for a given value of the RH gauge boson mass $M_{W_R}$ and the mixing parameter $|V_{\ell N}|$. This is shown in Fig.~\ref{fig3} for a typical choice of $M_{W_R}=3$ TeV, keeping in mind the current limit from direct collider searches which 
extend up to $M_{W_R}=2.5$ TeV~\cite{ATLAS-RR,CMS-RR}, and similar lower limits 
from estimates on the $K_L-K_S$ mixing~\cite{KL-KS}. We have only considered $\ell=\mu$ final state for our collider analysis since the heavy neutrino mixing to 
electrons is highly constrained from $0\nu\beta\beta$~\cite{Benes:2005hn}: $M_{W_R}^{-4}|\sum_i V^2_{eN_i}/M_{N_i}|<0.1~{\rm TeV}^{-5}$. Also we do not consider $\tau$ final states since the $\tau$-lepton identification at the LHC is rather complicated.  
We have shown the results 
for $\sqrt s=14$ TeV LHC and for two sample choices of the mixing: (a) $|V_{\ell N}|^2= 3\times 10^{-3}$, close to the current experimental limit on $|V_{\mu N}|^2$ for a TeV-scale heavy neutrino~\cite{delAguila:2008pw}, and (b) the vanilla seesaw expectation: $|V_{\ell N}|^2=\sqrt{\Delta m^2_{\rm atm}}/M_N$, $\Delta m^2_{\rm atm}$ being the atmospheric neutrino mass-squared difference which we take as $2.35\times 10^{-3}~{\rm eV}^2$~\cite{PDG}. 

The heavy neutrino signal cross section is given by 
\begin{eqnarray}
\sigma(pp\to N\ell^\pm \to \ell^\pm\ell^\pm jj)&=&\sigma_{\rm prod}(pp\to W_{L,R}\to N\ell^\pm)\nonumber\\
&&\times~ {\rm BR}(N\to \ell^\pm jj).
\label{eq:2}
\end{eqnarray}
 The parton-level production cross sections were generated for $\sqrt s=14$ TeV using {\tt CalcHEP}~\cite{calchep} with the {\tt CTEQ6L} parton distribution functions~\cite{cteq6l}. For the $LL$ and $RL$ modes, we have the 2-body decay $N\to \ell^\pm W$ followed by $W\to jj$, with the corresponding branching ratio
% in Eq.~(\ref{eq:2}) given by 
\begin{eqnarray}
{\rm BR}(N\to \ell^\pm jj) = \frac{\Gamma(N\to \ell^\pm W)}{\Gamma_N^{\rm tot}}\times {\rm BR}(W\to jj),
\label{eq:3}
\end{eqnarray}
where ${\rm BR}(W\to jj)=0.676$~\cite{PDG}. For the $RR$ mode, we have the 
three-body decay $N\to \ell^\pm W_R^*\to \ell^\pm jj$.  
The total decay width $\Gamma_N^{\rm tot}$ is the sum of partial 
widths to 2-body final states (when kinematically allowed): 
%$\ell^\pm W,\nu_\ell Z, \bar{\nu}_\ell Z, \nu_\ell h, \bar{\nu}_\ell h$:
%\begin{widetext}
\begin{eqnarray}
&&\Gamma(N\to \ell^\pm W) = \frac{g^2|V_{\ell N}|^2}{64\pi}\frac{M_N^3}{M_W^2}\left(1-\frac{M_W^2}{M_N^2}\right)^2\left(1+2\frac{M_W^2}{M_N^2}\right),\nonumber \\
&&\Gamma(N\to \nu_\ell Z,~\bar{\nu}_\ell Z) = \frac{g^2|V_{\ell N}|^2}{128\pi\cos^2\theta_W}\frac{M_N^3}{M_Z^2}\left(1-\frac{M_Z^2}{M_N^2}\right)^2\nonumber\\
&&\hspace{5cm}\times 
\left(1+2\frac{M_Z^2}{M_N^2}\right),\nonumber \\
&&\Gamma(N\to \nu_\ell h,~\bar\nu_\ell h)= \frac{g^2|V_{\ell N}|^2}{128\pi}\frac{M_N^3}{M_W^2}\left(1-\frac{M_h^2}{M_N^2}\right)^2, \nonumber
\end{eqnarray}
and 3-body final states (in the limit of massless final states, and assuming $W_R, Z_R$ highly off-shell):
\begin{eqnarray}
&&\Gamma(N\to \ell^\pm W_R^*\to \ell^\pm jj) \simeq \frac{3g_R^4}{2048\pi^3}\frac{M_N^5}{M_{W_R}^4}, \nonumber \\ %\label{eq:7}\\
&&\Gamma(N\to \nu_\ell (\bar \nu_\ell) Z_R^*\to \nu_\ell (\bar\nu_\ell) jj) \simeq \frac{3g_R^4}{4096\pi^3}\frac{\cos^8\theta_W}{\cos^2{2\theta_W}}\frac{M_N^5}{M_{Z_R}^4}. \nonumber %\label{eq:8}
\end{eqnarray}
%\end{widetext}

For numerical purposes, we use $m_h=125$ GeV, $g_L=g_R$ for the weak gauge couplings, and the relation $M_{Z_R}/M_{W_R}=\cos\theta_W/\sqrt{\cos{2\theta_W}}$ (where $\theta_W$ is the Weinberg angle), assuming that the LR-symmetry is broken by an $SU(2)_R$ triplet Higgs vacuum expectation value.  We neglect the contribution of the 3-body decay modes of $N$ mediated by the $SU(2)_L$ triplet Higgs fields to its total width, since it not only involves the heavy-light mixing $|V_{\ell N}|^2$ 
(as the $N\to \ell W$ mode) but is further suppressed by the factor $M^5_N/M^4_{\Delta_{L}}$ (assuming $M_{\Delta_L}\gg M_N$) as well as the 3-body phase 
space.

It is clear from Fig.~\ref{fig3} that for small heavy-light neutrino mixing 
(right panel), 
the $RR$ mode is dominant for a TeV-scale $W_R$, while for large mixing 
(left panel), 
the $RL$ mode is dominant over both $LL$ and $RR$ modes over a wide range of RH neutrino masses relevant for their collider searches. Hence for consistency 
the $RL$ mode must also be taken into account in the collider analysis of heavy 
neutrinos in a TeV-scale LR-model.    

%%%%%%%%%%%%
\section{Improved Collider Limits on the Heavy-light Neutrino Mixing}
%%%%%%%%%%%%
As an immediate implication of our results shown above, we can derive 
improved collider limits on the left-right neutrino mixing compared to the existing limits~\cite{CMS-LL, ATLAS-LL} obtained from $\sqrt s=7$ TeV LHC data 
assuming the inclusive signal cross section for the $LL$ mode alone. 
For the range of mixing parameter being constrained here, the $RL$ contribution 
is in general dominant, especially for higher $M_N$ (cf. Fig.~\ref{fig3}) and 
the total ($LL+RL$) inclusive 
cross section is larger thus yielding a stronger limit on the mixing parameter. %In Fig.~\ref{fig4}, we show the ratio of the signal cross section for the RL and LL modes for both $\sqrt s=7$ and 14 TeV LHC. The LL cross section is normalized to $\sigma_{\rm LL}/|V_{\ell N}|^2\equiv \tilde{\sigma}_{LL}$ for comparison with the RL mode which is independent of the mixing. % but depends on the $W_R$ mass. 
%Note that the ratio of 
%signal cross sections in this case will be the same as the production cross section ratio since the heavy neutrino decay rate and its branching fraction are the same for both channels. 
We do not include the $RR$ contribution to the signal cross section since it 
is sub-dominant for the range of 
$|V_{\ell N}|$ considered here, and moreover, this channel will have a 
significantly smaller efficiency after applying the selection cuts designed 
for $LL$ mode~\cite{CMS-LL, ATLAS-LL} (also valid for $RL$ mode)-- in particular, the requirement of the dijet 
invariant mass $m_{jj}$ close to $M_W$.   
%\begin{figure}[h!]
%\centering
%\includegraphics[width=4cm]{fig4a.pdf}
%\includegraphics[width=4cm]{fig4b.pdf}
%\caption{The ratio of the signal cross sections for RL and LL modes at $\sqrt s=7$ and 14 TeV LHC.} 
%The LL cross section is normalized to $\sigma_{\rm LL}/|V_{\ell N}|^2\equiv \tilde{\sigma}_{LL}$ for comparison with $\sigma_{\rm RL}$.}
%\label{fig4}
%\end{figure}
%We note from Fig.~\ref{fig3} that the total (LL+RL) 
%inclusive cross 
%section could be much higher than the LL cross section alone, especially for 
%higher $M_N$. 

Thus, given an experimentally observed limit on the signal cross section $\sigma_{\rm expt}$, we can infer the following: (i) the $(M_N,M_{W_R})$ plane for which $\sigma_{RL}\geq \sigma_{\rm expt}$ is ruled out, thus providing a complementary probe of this 
parameter space which is currently probed at the LHC only in the $RR$ mode~\cite{ATLAS-RR, CMS-RR}; (ii) for $\sigma_{RL}<\tilde\sigma_{LL}<\sigma_{\rm expt}$ where $\tilde{\sigma}_{LL}\equiv \sigma_{LL}/|V_{\ell N}|^2$ is the normalized $LL$ cross section, an improved limit on the mixing parameter can be derived:
\begin{eqnarray}
|V_{\ell N}|^2 < \frac{\sigma_{\rm expt}-\sigma_{RL}}{\tilde{\sigma}_{LL}}
%\tilde\sigma_{\rm LL}|V_{\ell N}|^2+\sigma_{\rm RL} < \sigma_{\rm expt}
\label{eq:9}
\end{eqnarray}
which is obviously stronger than that derived assuming $\sigma_{RL}=0$. 
Using the observed cross section limit for $\sqrt s=7$ TeV from the ATLAS analysis~\cite{ATLAS-LL}, we find the improvement in the upper limit 
on $|V_{\ell N}|^2$ taking into account the combined ($LL+RL$) mode in the minimal LR model with $M_{W_R}=2.5$ TeV to be about 10\% for $M_N=300$ GeV, and 
somewhat lower for decreasing (increasing) $M_N~(M_{W_R})$.
%shown in Table~\ref{tab1}. The current ATLAS limits derived from the LL mode alone are also shown for comparison. Thus we find an improvement of up to 
However, we expect 
it to be much more prominent for higher values of $M_N$ and/or at $\sqrt s=14$ TeV LHC due to the enhanced $RL$ cross 
section as shown in Fig.~\ref{fig3}. For illustration, assuming the expected 
upper limit on the signal cross section at $\sqrt s=14$ TeV LHC to be 
smaller than the observed limit at $\sqrt s=7$ TeV, we obtain conservative 
upper limits on the mixing parameter as shown in Table~\ref{tab2}. 
The low $M_{W_R}$ points denoted by a * predict cross sections larger than 
our assumed experimental limit, and hence, can be ruled out in case of no 
positive signal. On the other hand, for the allowed region, the improvement in 
the limit on mixing could be as large as 60\%.     
%we obtain using Eq.~(\ref{eq:9}) the upper limit 
%on $|V_{\ell N}|^2$ to be 0.1614 (0.0095) at $M_{W_R}=2.5$ TeV and 
%$M_N=300~(100)$ GeV, compared to the existing limit of 0.18 (0.01)~\cite{ATLAS-LL}. We expect this improvement to be much more prominent for higher values of 
%
%For illustration,  we obtain the following conservative upper limit on $|V_{\ell N}|^2$ for $M_N=100$ GeV: $1.2\times 10^{-3}$ from the LL channel only, compared to $4.6~(9.3)\times 10^{-4}$ from the (LL+RL) channel for $M_{W_R}=3~(3.5)$ TeV.

\begin{table}[h!]
\begin{center}
\begin{tabular}{c|c||c|c|c}%||c|c|c}
\hline\hline
Mode & $M_{W_R}$ & \multicolumn{3}{|c}{Upper limit on 
$|V_{\ell N}|^2$ for $\sqrt s=14$ TeV LHC} 
% & \multicolumn{3}{||c||}{$\sqrt s=14$ TeV} 
\\ \cline{3-5}
& (TeV) & $M_N=100$ GeV & $M_N=200$ GeV & $M_N=300$ GeV \\ \hline
 & 2.5 & * & * & * \\ \cline{2-5} 
$LL$ & 3 & 0.0005 & * & * \\ \cline{2-5}
+ & 3.5 & 0.0009 & * & * \\ \cline{2-5}
$RL$ & 4 & 0.0011 & 0.0013 & 0.0042 \\ \cline{2-5}
& 5 & 0.0012 & 0.0026 & 0.0092 \\ \hline 
\multicolumn{2}{c||}{$LL$} & 0.0012 & 0.0029 & 0.0102\\ 
\hline\hline
\end{tabular}
\end{center}
\caption{Projected upper limits on the heavy-light neutrino mixing in the 
minimal LR model for $\sqrt s=14$ TeV LHC. The * points predict a cross 
section larger than our expected $\sigma_{\rm expt}$.}
\label{tab2}
\end{table} 
 
 %%%%%%%%%%%%%%%%%%%
\section{Post-Discovery Distinction}
\begin{figure}[htb]
\centering
\includegraphics[width=4cm]{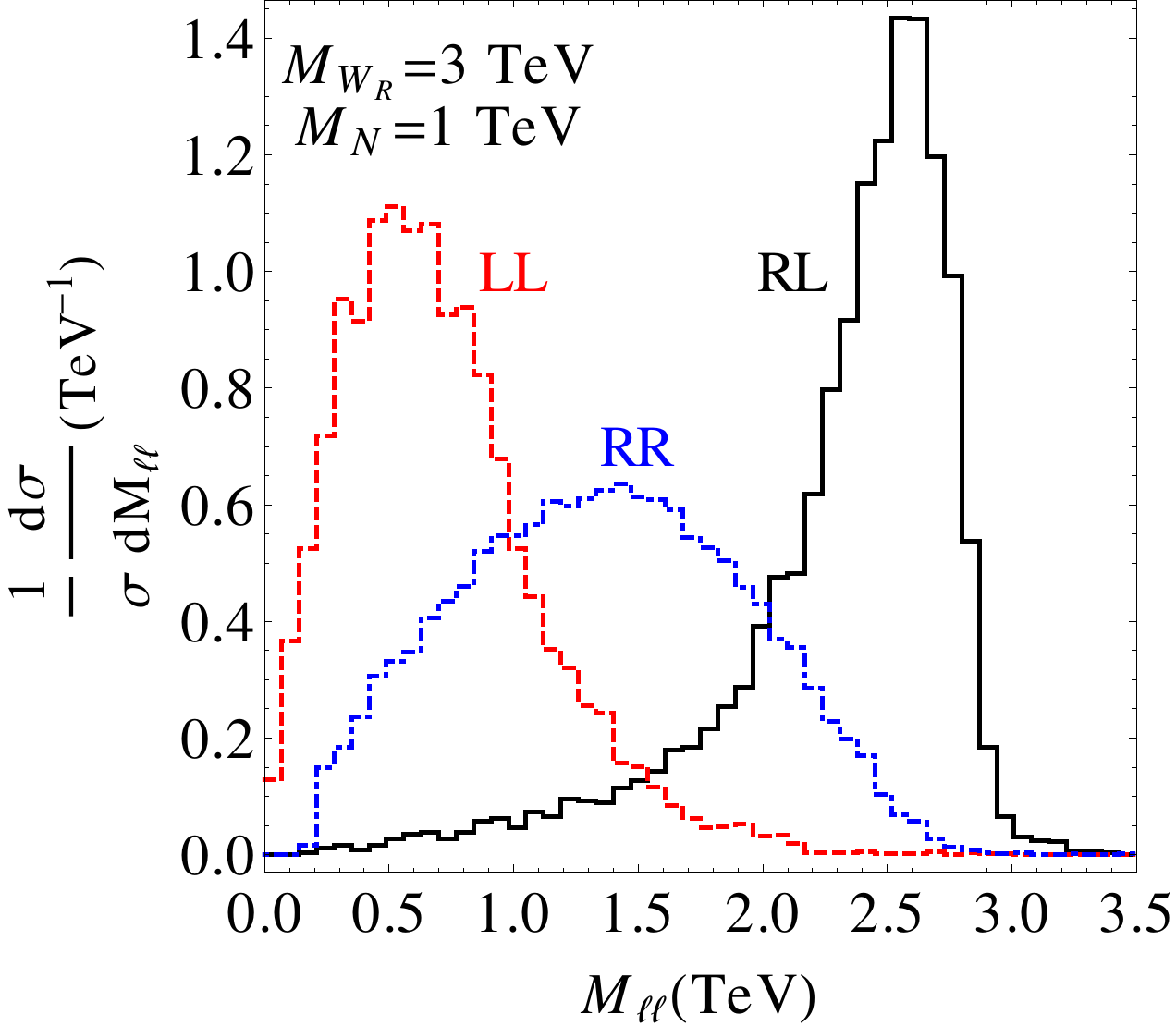}
\hspace{0.0cm}
\includegraphics[width=4cm]{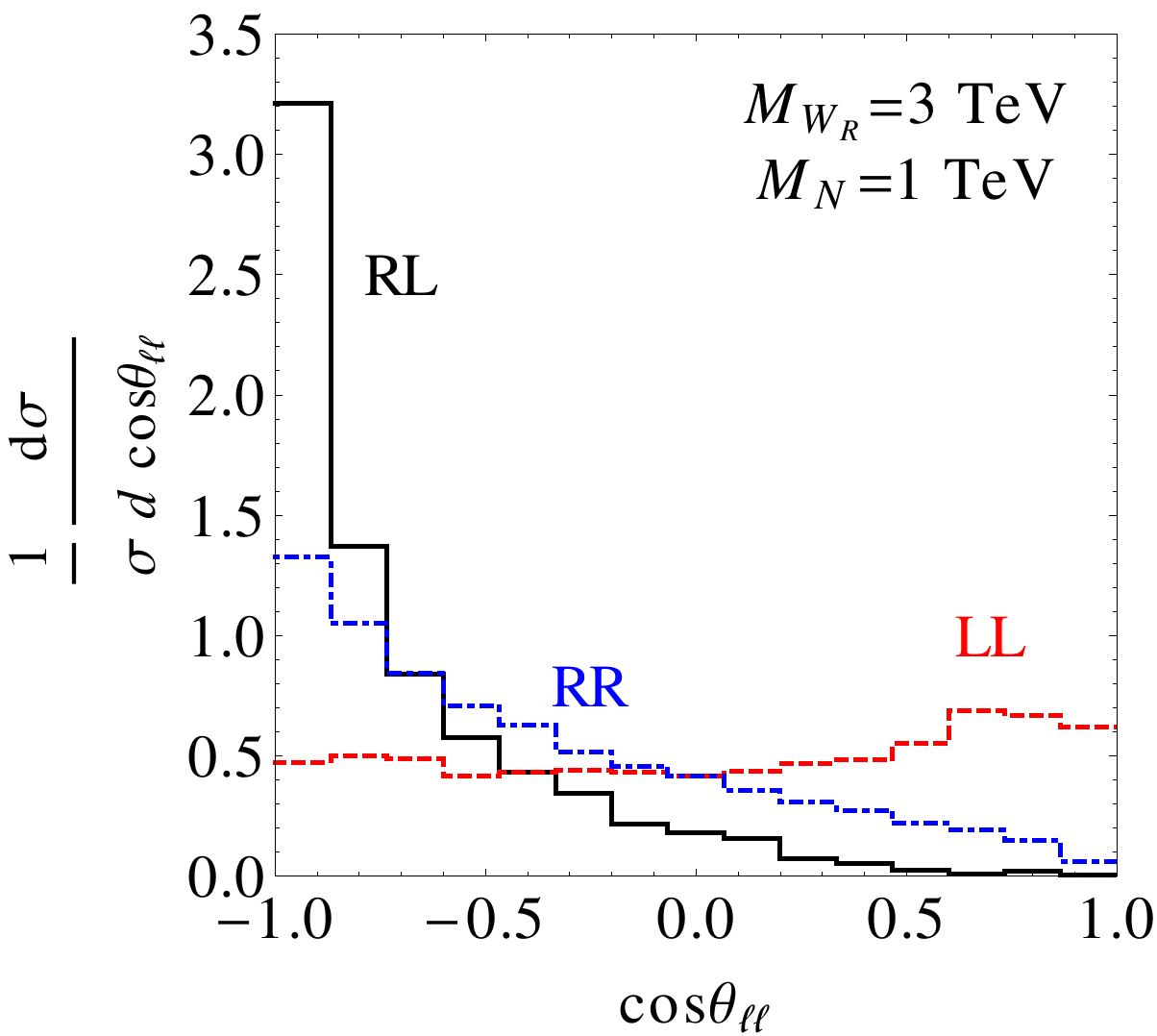}
\caption{The invariant mass distribution and the angular 
correlation of the final state leptons for the $LL$, $RL$ and $RR$ modes shown in Fig.~\ref{fig1}. For comparison, all distributions have been normalized to unity.}
\label{fig5}
\end{figure}

%%%%%%%%%%%%%%%%
Here we propose a possible distinction between the various contributions shown in Fig.~\ref{fig2} %which can be used to prove the existence of a low-scale LR symmetry 
%if a same-sign dilepton plus two jet signal with no missing energy is observed  at the LHC. 
by considering two kinematic variables, namely, the dilepton invariant mass distribution and angular correlation between the charged leptons. For a 
realistic collider simulation,   
%Our simulation results are shown 
%in Fig.~\ref{fig5}. %, after implementing the realistic cuts from existing  
%experimental analyses.% and taking into account the detector effects. 
%The parton-level signal events are calculated using {\tt CalcHEP}~\cite{calchep}, which are then fed into {\tt PYTHIA}~\cite{pythia} to implement initial and final state radiation of quarks and gluons, multiple interactions, decay, hadronization, fragmentation and jet formation. Finally, a fast detector simulation is performed using {\tt PGS}~\cite{PGS}. 
the parton-level signal events generated by {\tt CalcHEP}~\cite{calchep} are fed into {\tt PYTHIA}~\cite{pythia} and {\tt PGS4}~\cite{PGS} to implement parton showering, hadronization and detector effects.
We have used an anti-$k_T$ jet algorithm with jet cone size parameter $R=0.4$. Apart from the basic selection criteria of two same-sign muons and two light 
jets, we have implemented the following selection cuts for both $LL$ and $RL$ 
modes following the latest ATLAS analysis~\cite{ATLAS-LL}: $p_T^j>20$ GeV, $p_T^\ell>20$ GeV, $p_T^{\ell,\rm leading}>25$ GeV, $|\eta(j)|<2.8$, $|\eta(\ell)|<2.5$, $\mET<35$ GeV and $m_{jj}\in[55,120]$ GeV. For the $RR$ mode, 
we have implemented the cuts following the latest CMS analysis~\cite{CMS-RR}: 
$M_{\ell\ell jj}>600$ GeV, $M_{\ell\ell}>200$ GeV, $p_T^j>40$ GeV, $p_T^\ell>40$ GeV, $p_T^{\ell,{\rm leading}}>60$ GeV, $|\eta(j)|<3.0$ and 
$|\eta(\ell)|<2.5$. For comparison, all distributions have been normalized to unity after applying the cuts. 
%The parton-level cross sections are normalized to give the 
%same number of events for each mode before applying the cuts. 
The simulation results for an illustrative case with $|V_{\ell N}|^2=0.003$, $M_{W_R}=3$ TeV and $M_N=1$ TeV are shown in Fig.~\ref{fig5}. 
It is clear that the dilepton invariant mass distribution (left panel) 
is a good kinematic variable for distinction between the $LL$, $RL$ and $RR$ modes at the LHC. The angular correlation between the two leptons (right panel) 
is another good variable 
to distinguish the $RL$ and $RR$ case from the $LL$ case due to different helicity correlations. 
%The angular distribution of the charged lepton originating from the heavy neutrino decay in the $N$ 
%rest frame with respect to the $N$ moving direction in the partonic center of mass frame could also be used to distinguish these two cases~\cite{Han:2012vk}, but after implementing all the selection criteria described above, we did not 
%find our RL and LL cases to be well separated for this variable. 

%%%%%%%%%%%%%%%%%%%%%%%%
\section {Conclusion}
%%%%%%%%%%%%%%%%%%%%%%
In summary, we have pointed out a new contribution to the smoking gun collider signals of a TeV scale left-right seesaw model i.e. $\ell^\pm\ell^\pm jj$ coming from the heavy-light neutrino mixing contribution (called $RL$ in the text) which can dominate over the usually discussed $LL$ and $RR$ contributions.  Probing this contribution can provide crucial information on the detailed nature of the seesaw mechanism and supplement searches for this effect using violations of unitarity of the PMNS matrix. This will provide extremely important information regarding the detailed nature of left-right TeV scale seesaw models.  We emphasize the importance of this channel for heavy Majorana neutrino searches in the hope that  this will be taken into account in the future experimental analyses, along with the usual $LL$ and $RR$ channels. We show how taking into account this $RL$ contribution can improve the collider limits on the left-right neutrino mixing in certain parameter domains of the seesaw matrix, with the improvement becoming more prominent as we go to higher heavy neutrino masses and higher center of mass energy at the LHC. We also propose a simple way to distinguish the different contributions and to identify the dominant channel by analyzing  the invariant mass distribution and angular correlation of the two same-sign leptons. Should a same-sign dilepton plus two jets with no missing energy signal be observed at the LHC, this will help us in determining the existence of a TeV-scale LR-symmetry as well as the structure of the seesaw matrix. 

\noindent{\bf Acknowledgments}: 
C-Y.C and P.S.B.D. thank Ian Lewis and Apostolos Pilaftsis for helpful 
discussions. 
P.S.B.D. also thanks the 
High Energy Theory group at BNL for hospitality 
where part of this work was carried out. 
The work of C-Y.C. is supported by the US 
Department of Energy under Grant DE-AC02-98CH10886, P.S.B.D. is supported by 
the Lancaster-Manchester-Sheffield Consortium for Fundamental Physics under 
STFC grant ST/J000418/1, and R.N.M. is supported by National Science 
Foundation grant No. PHY-0968854.


\begin{thebibliography}{99}
\bibitem{type1} P. Minkowski, Phys. Lett. B {\bf 67}, 421 (1977);  
T. Yanagida, {\it Workshop on unified theories and baryon number in the universe}, eds. A. Sawada and A. Sugamoto, KEK, Tsukuba (1979); 
%S. Glashow, Quarks and lep- tons,Carg`ese1979,ed.M.L ́evy(Plenum,NY,1980); 
M. Gell-Mann, P. Ramond and R. Slansky, {\it Supergravity}, ed. P. Van Niewenhuizen and D. Freeman, North Holland, Amsterdam (1980); 
R. N. Mohapatra and G. Senjanovi\'{c}, Phys. Rev. Lett. {\bf 44}, 912 (1980).

\bibitem{valle} J.~Schechter and J.~W.~F.~Valle,
 %``Neutrino Masses in SU(2) x U(1) Theories,''
  Phys.\ Rev.\ D {\bf 22}, 2227 (1980); 
  %``Neutrino Decay and Spontaneous Violation of Lepton Number,''
  Phys.\ Rev.\ D {\bf 25}, 774 (1982).

\bibitem{0v2breview} For reviews, see e.g., W. Rodejohann, Int. J. Mod. Phys. E {\bf 20}, 1833 (2011); %[arXiv:1106.1334 [hep-ph]]; 
J. D. Vergados, H. Ejiri and F. Simkovic, Rept. Prog. Phys. {\bf 75}, 106301 (2012).% [arXiv:1205.0649 [hep-ph]];

 
  \bibitem{unitarity} 
% H.~Zhang and Z.~-z.~Xing,
  %``Leptonic unitarity triangles in matter,''
 % Eur.\ Phys.\ J.\ C {\bf 41}, 143 (2005); 
S.~Antusch, C.~Biggio, E.~Fernandez-Martinez, M.~B.~Gavela and J.~Lopez-Pavon,
  %``Unitarity of the Leptonic Mixing Matrix,''
  JHEP {\bf 0610}, 084 (2006); 
A.~Abada, C.~Biggio, F.~Bonnet, M.~B.~Gavela and T.~Hambye,
  %``Low energy effects of neutrino masses,''
  JHEP {\bf 0712}, 061 (2007); 
%  [arXiv:0707.4058 [hep-ph]].
M.~Malinsky, T.~Ohlsson and H.~Zhang,
  %``Non-unitarity effects in a realistic low-scale seesaw model,''
  Phys.\ Rev.\ D {\bf 79}, 073009 (2009); 
M.~Malinsky, T.~Ohlsson, Z.~-z.~Xing and H.~Zhang,
  %``Non-unitary neutrino mixing and CP violation in the minimal inverse seesaw model,''
  Phys.\ Lett.\ B {\bf 679}, 242 (2009); 
%  [arXiv:0905.2889 [hep-ph]].  
P.~S.~B.~Dev and R.~N.~Mohapatra,
  %``TeV Scale Inverse Seesaw in SO(10) and Leptonic Non-Unitarity Effects,''
  Phys.\ Rev.\ D {\bf 81}, 013001 (2010). 
%  [arXiv:0910.3924 [hep-ph]].

\bibitem{Drewes:2013gca} 
  For a review, see e.g., M.~Drewes,
  %``The Phenomenology of Right Handed Neutrinos,''
  arXiv:1303.6912 [hep-ph].
 

\bibitem{LR} J.C. Pati and A. Salam, Phys. Rev. D {\bf 10}, 275 (1974); 
R. N. Mohapatra and J. C. Pati, Phys. Rev. D {\bf 11} 2558 (1975); 
R. N. Mohapatra and G. Senjanovi\'{c}, Phys. Rev. D {\bf 12} 1502 (1975). 

\bibitem{goran} G. Senjanovi\'{c}, Int.\ J.\ Mod.\ Phys.\ A {\bf 26}, 1469 (2011). 

\bibitem{theory-LL} A.~Datta, M.~Guchait and A.~Pilaftsis,
  %``Probing lepton number violation via majorana neutrinos at hadron supercolliders,''
  Phys.\ Rev.\ D {\bf 50}, 3195 (1994); 
F.~M.~L.~Almeida, Jr. {\it et al.}, %Y.~D.~A.~Coutinho, J.~A.~Martins Simoes and M.~A.~B.~do Vale,
  %``On a signature for heavy Majorana neutrinos in hadronic collisions,''
  Phys.\ Rev.\ D {\bf 62}, 075004 (2000); 
T. Han and B. Zhang, Phys. Rev. Lett. {\bf 97}, 171804 (2006); 
F. del Aguila, J. A. Aguilar-Saavedra and R. Pittau, JHEP {\bf 10}, 047 (2009); 
F.~del Aguila and J.~A.~Aguilar-Saavedra,
  %``Distinguishing seesaw models at LHC with multi-lepton signals,''
  Nucl.\ Phys.\ B {\bf 813}, 22 (2009).

\bibitem{Planck} P. A. R. Ade {\it et al.} [Planck Collaboration], arXiv:1303.5076 [astro-ph.CO]. 

\bibitem{cancel} A. Pilaftsis, Z. Phys. C {\bf 55}, 275 (1992); 
J.~Gluza,
  %``On teraelectronvolt Majorana neutrinos,''
  Acta Phys.\ Polon.\ B {\bf 33}, 1735 (2002);  
J. Kersten and A. Y. Smirnov, Phys. Rev. D {\bf 76}, 073005 (2007); A.~de Gouvea, arXiv:0706.1732 [hep-ph];
  %``GeV seesaw, accidentally small neutrino masses, and Higgs decays to neutrinos,'';
Z.~-z.~Xing,
  %``Naturalness and Testability of TeV Seesaw Mechanisms,''
  Prog.\ Theor.\ Phys.\ Suppl.\  {\bf 180}, 112 (2009); 
X.~-G.~He, S.~Oh, J.~Tandean and C.~-C.~Wen,
  %``Large Mixing of Light and Heavy Neutrinos in Seesaw Models and the LHC,''
  Phys.\ Rev.\ D {\bf 80}, 073012 (2009); 
 A.~Ibarra, E.~Molinaro and S.~T.~Petcov,
  %``TeV Scale See-Saw Mechanisms of Neutrino Mass Generation, the Majorana Nature of the Heavy Singlet Neutrinos and $(\beta\beta)_{0\nu}$-Decay,''
  JHEP {\bf 1009}, 108 (2010); 
 % M.~Mitra, G.~Senjanovi\'{c} and F.~Vissani,
  %``Neutrinoless Double Beta Decay and Heavy Sterile Neutrinos,''
 % Nucl.\ Phys.\ B {\bf 856}, 26 (2012); 
N.~Haba, T.~Horita, K.~Kaneta and Y.~Mimura,
  %``TeV-scale seesaw with non-negligible left-right neutrino mixings,''
  arXiv:1110.2252 [hep-ph].

\bibitem{Mitra:2011qr} 
 M.~Mitra, G.~Senjanovic and F.~Vissani,
  %``Neutrinoless Double Beta Decay and Heavy Sterile Neutrinos,''
  Nucl.\ Phys.\ B {\bf 856}, 26 (2012). 
 % [arXiv:1108.0004 [hep-ph]]; 
 

\bibitem{Blennow:2010th} 
  M.~Blennow, E.~Fernandez-Martinez, J.~Lopez-Pavon and J.~Menendez,
  %``Neutrinoless double beta decay in seesaw models,''
  JHEP {\bf 1007}, 096 (2010). 
  %[arXiv:1005.3240 [hep-ph]]. 

\bibitem{Ibarra:2011xn} 
  A.~Ibarra, E.~Molinaro and S.~T.~Petcov,
  %``Low Energy Signatures of the TeV Scale See-Saw Mechanism,''
  Phys.\ Rev.\ D {\bf 84}, 013005 (2011); 
 % [arXiv:1103.6217 [hep-ph]]; 
%\bibitem{LopezPavon:2012zg} 
J.~Lopez-Pavon, S.~Pascoli and C.~-f.~Wong,
  %``Can heavy neutrinos dominate neutrinoless double beta decay?,''
  arXiv:1209.5342 [hep-ph].

\bibitem{Tello:2010am} 
  V.~Tello, M.~Nemevsek, F.~Nesti, G.~Senjanovi\'{c} and F.~Vissani,
  %``Left-Right Symmetry: from LHC to Neutrinoless Double Beta Decay,''
  Phys.\ Rev.\ Lett.\  {\bf 106}, 151801 (2011). 
%  [arXiv:1011.3522 [hep-ph]].

\bibitem{MS} R. N. Mohapatra and G. Senjanovi\'{c}, in Ref.~\cite{type1}; 
Phys. Rev. D {\bf 23}, 165 (1981). 
%\bibitem{Mohapatra:1981pj} 
% Riazuddin, R.~E.~Marshak, R.~N.~Mohapatra,
  %``Majorana Neutrinos And Low-energy Tests Of Electroweak Models,''
  %Phys.\ Rev.\ D {\bf 24}, 1310 (1981). 

\bibitem{KS} W.-Y. Keung and G. Senjanovi\'{c}, Phys. Rev. Lett. {\bf 50}, 1427 (1983).

\bibitem{theory-RR}  A.~Datta, M.~Guchait and D.~P.~Roy,
  %``Prospect of heavy right-handed neutrino search at SSC / CERN LHC energies,''
  Phys.\ Rev.\ D {\bf 47}, 961 (1993); 
%  [hep-ph/9208228].
A.~Ferrari {\it et al.}, 
%J.~Collot, M-L.~Andrieux, B.~Belhorma, P.~de Saintignon, J-Y.~Hostachy, P.~.Martin and M.~Wielers,
  %``Sensitivity study for new gauge bosons and right-handed Majorana neutrinos in $p p$ collisions at $s$ = 14-TeV,''
  Phys.\ Rev.\ D {\bf 62}, 013001 (2000);
S. N. Gninenko, M. M. Kirsanov, N. V. Krasnikov and V. A. Matveev, 
Phys. Atom. Nucl. {\bf 70}, 441 (2007); 
%A.~Maiezza, M.~Nemevsek, F.~Nesti and G.~Senjanovic,
  %``Left-Right Symmetry at LHC,''
  %Phys.\ Rev.\ D {\bf 82}, 055022 (2010)
  %[arXiv:1005.5160 [hep-ph]]; 
  M.~Nemevsek, F.~Nesti, G.~Senjanovi\'{c} and Y.~Zhang,
  %``First Limits on Left-Right Symmetry Scale from LHC Data,''
  Phys.\ Rev.\ D {\bf 83}, 115014 (2011); 
J.~Chakrabortty, J.~Gluza, R.~Sevillano and R.~Szafron,
  %``Left-Right Symmetry at LHC and Precise 1-Loop Low Energy Data,''
  JHEP {\bf 1207}, 038 (2012);
%  [arXiv:1204.0736 [hep-ph]]; 
S.~P.~Das, F.~F.~Deppisch, O.~Kittel and J.~W.~F.~Valle,
  %``Heavy Neutrinos and Lepton Flavour Violation in Left-Right Symmetric Models at the LHC,''
  Phys.\ Rev.\ D {\bf 86}, 055006 (2012); 
J.~A.~Aguilar-Saavedra and F.~R.~Joaquim,
  %``Measuring heavy neutrino couplings at the LHC,''
  Phys.\ Rev.\ D {\bf 86}, 073005 (2012). 
 % [arXiv:1207.4193 [hep-ph]].

\bibitem{rnm} R.~N.~Mohapatra,
  %``Limits On The Mass Of The Right-handed Majorana Neutrino,''
  Phys.\ Rev.\ D {\bf 34}, 909 (1986).


  
%  \bibitem{delaguila} F.~del Aguila and J.~A.~Aguilar-Saavedra,
  %``Distinguishing seesaw models at LHC with multi-lepton signals,''
%  Nucl.\ Phys.\ B {\bf 813}, 22 (2009)
%  [arXiv:0808.2468 [hep-ph]].


\bibitem{dlm} P. S. B. Dev, C. Lee and R. N. Mohapatra, work in progress.


\bibitem{CMS-LL} CMS Collaboration, Phys. Lett. {\bf B 717}, 109 (2012). 

\bibitem{ATLAS-LL} ATLAS Collaboration, ATLAS-CONF-2012-139.  

\bibitem{Atre:2009rg} 
  A.~Atre, T.~Han, S.~Pascoli and B.~Zhang,
  %``The Search for Heavy Majorana Neutrinos,''
  JHEP {\bf 0905}, 030 (2009).


%\bibitem{Tommasini:1995ii}
%D.~Tommasini, G.~Barenboim, J.~Bernabeu and C.~Jarlskog,
  %``Nondecoupling of heavy neutrinos and lepton flavor violation,''
%  Nucl.\ Phys.\ B {\bf 444}, 451 (1995)
%  [hep-ph/9503228].

\bibitem{ATLAS-RR} ATLAS Collaboration, Eur. Phys. J. {\bf C 72}, 2056 (2012). 

\bibitem{CMS-RR} CMS Collaboration, CMS-PAS-EXO-12-017.


\bibitem{Chen:2011hc} 
  C.~-Y.~Chen and P.~S.~B.~Dev,
  %``Multi-Lepton Collider Signatures of Heavy Dirac and Majorana Neutrinos,''
  Phys.\ Rev.\ D {\bf 85}, 093018 (2012).

\bibitem{Han:2012vk} 
  T.~Han, I.~Lewis, R.~Ruiz and Z.~-g.~Si,
  %``Lepton Number Violation and W' Chiral Couplings at the LHC,''
  Phys.\ Rev.\ D {\bf 87}, 035011 (2013).

\bibitem{Hirsch:1996qw} 
 M.~Hirsch, H.~V.~Klapdor-Kleingrothaus and O.~Panella,
  %``Double beta decay in left-right symmetric models,''
  Phys.\ Lett.\ B {\bf 374}, 7 (1996). 
%  [hep-ph/9602306].


\bibitem{Nemevsek:2012iq} 
 M.~Nemevsek, G.~Senjanovi\'{c} and V.~Tello,
  %``Left-Right Symmetry: from Majorana to Dirac,''
 Phys. Rev. Lett. {\bf 110}, 151802 (2013);  
%[arXiv:1211.2837 [hep-ph]].
%J.~Chakrabortty, H.~Z.~Devi, S.~Goswami and S.~Patra,
  %``Neutrinoless double-$\beta$ decay in TeV scale Left-Right symmetric models,''
%  JHEP {\bf 1208}, 008 (2012)
%  [arXiv:1204.2527 [hep-ph]];
J.~Barry and W.~Rodejohann,
  %``Lepton number and lepton flavour violation in left-right symmetric theories,''
  arXiv:1303.6324 [hep-ph].


\bibitem{KL-KS} G.~Beall, M.~Bander and A.~Soni,
  %``Constraint on the Mass Scale of a Left-Right Symmetric Electroweak Theory from the K(L) K(S) Mass Difference,''
  Phys.\ Rev.\ Lett.\  {\bf 48}, 848 (1982); 
H. An, X. Ji, R. N. Mohapatra, and Y. Zhang, Nucl. Phys. {\bf B 802}, 247 
(2008); %[arXiv:0712.4218 [hep-ph]];
A.~Maiezza, M.~Nemevsek, F.~Nesti and G.~Senjanovi\'{c},
  %``Left-Right Symmetry at LHC,''
  Phys.\ Rev.\ D {\bf 82}, 055022 (2010). 
 % [arXiv:1005.5160 [hep-ph]].
 

\bibitem{Benes:2005hn} 
  P.~Benes, A.~Faessler, F.~Simkovic and S.~Kovalenko,
  %``Sterile neutrinos in neutrinoless double beta decay,''
  Phys.\ Rev.\ D {\bf 71}, 077901 (2005); 
 P.~S.~B.~Dev, S.~Goswami, M.~Mitra and W.~Rodejohann,
  %``Constraining Neutrino Mass from Neutrinoless Double Beta Decay,''
  arXiv:1305.0056 [hep-ph].

\bibitem{delAguila:2008pw} 
  F.~del Aguila, J.~de Blas and M.~Perez-Victoria,
  %``Effects of new leptons in Electroweak Precision Data,''
  Phys.\ Rev.\ D {\bf 78}, 013010 (2008).
  
  
\bibitem{PDG} J. Beringer {\it et al.} (Particle Data Group), Phys. Rev. D {\bf 86}, 010001 (2012). 
 

\bibitem{calchep} A. Pukhov {\it et al.}, hep-ph/9908288; A. Pukhov, hep-ph/0412191. 

\bibitem{cteq6l} J. Pumplin {\it et al.}, JHEP {\bf 07}, 012 (2002). 



\bibitem{pythia} T. Sjostrand, S. Mrenna, and P. Z. Skands, JHEP {\bf 05}, 026 (2006). 

\bibitem{PGS} J. Conway, {\tt http://www.physics.ucdavis.edu/\textasciitilde conway/}\\
{\tt research/software/pgs/pgs4-general.htm}. 

\end{thebibliography}
\end{document}